\documentclass[]{aastex701}
\usepackage{amsmath}

\begin{document}

\title{Spatiotemporal Properties of Compressible Magnetohydrodynamic Turbulence from Space Plasma}

\author[orcid=0000-0003-4268-7763]{Siqi Zhao}
\affiliation{Institut für Physik und Astronomie, Universität Potsdam, 14476, Potsdam, Germany}
\affiliation{Deutsches Elektronen-Synchrotron DESY, Platanenallee 6, 15738, Zeuthen, Germany}
\affiliation{Institute of Science and Technology for Deep Space Exploration, Nanjing University, China}
\email{siqi.zhao@desy.de}  

\author[orcid=0000-0003-2560-8066]{Huirong Yan} 
\affiliation{Institut für Physik und Astronomie, Universität Potsdam, 14476, Potsdam, Germany}
\affiliation{Deutsches Elektronen-Synchrotron DESY, Platanenallee 6, 15738, Zeuthen, Germany}
\email[show]{huirong.yan@desy.de}

\author[orcid=0000-0003-1778-4289]{Terry Z. Liu}
\affiliation{Shandong Key Laboratory of Space Environment and Exploration Technology, Institute of Space Sciences, School of Space Science and Technology, Shandong University, Shandong, China.}
\email[show]{terryliuzixu@sdu.edu.cn}  

\author[orcid=0000-0001-7205-2449]{Chuanpeng Hou}
\affiliation{Institut für Physik und Astronomie, Universität Potsdam, 14476, Potsdam, Germany}
\email{}  

\author[0000-0003-1683-9153]{Ka Ho Yuen}
\affiliation{Institute of Science and Technology for Deep Space Exploration, Nanjing University, China}
\email{kyuen@nju.edu.cn}

\correspondingauthor{Huirong Yan; Terry Z. Liu}

\begin{abstract}
It is well established that Alfvénic turbulence undergoes a weak-to-strong transition as energy cascades to smaller scales, but whether this occurs in compressible magnetohydrodynamic (MHD) turbulence remains unclear. Using multi-spacecraft measurements from Cluster and an improved polarization-based mode-decomposition technique, we obtain spatiotemporal power spectra of Earth’s magnetosheath turbulence and quantify nonlinear frequency broadening of the three MHD eigenmodes. We find that slow modes transition from wave-like peaks to strongly turbulent, frequency-broadened behavior as energy cascades to smaller scales, whereas fast modes remain weakly turbulent with narrow spectral peaks near their eigenfrequencies. These findings provide the first observational characterization of mode-dependent spatiotemporal dynamics in compressible MHD turbulence, with implications for energetic particle transport, plasma heating, and solar wind–magnetosphere coupling.
\end{abstract}

\section{Introduction} 
Plasma turbulence is ubiquitous and plays a fundamental role in astrophysical and space plasmas, spanning environments from galaxy clusters and the interstellar medium to accretion disks and the heliosphere \citep{Zank1992, Armstrong1995, Brunetti2007, Lazarian2008, Crutcher2012, Bruno2013, Yan2022, Zhao2021,Zhao2025_si, Zhao2026}. Turbulent processes transfer energy across a wide range of spatial and temporal scales, typically characterized by broadband spectra in both frequency and wavenumber space \citep{Kolmogorov1941, Iroshnikov}. Previous simulations \citep{Verdini2012, Meyrand2016, Makwana2020} and observations \citep{Zhao2024a} have demonstrated a weak-to-strong transition in incompressible Alfvénic turbulence as energy cascades from large to small scales. However, the spatiotemporal (frequency–wavenumber) properties of compressible MHD turbulence involving all eigenmodes, which encode the strength of nonlinear interactions, remain difficult to characterize observationally. Consequently, whether a similar weak-to-strong transition occurs in compressible turbulence remains elusive. Understanding this self-organized process is essential for elucidating the dynamics in compressible turbulence.

In a homogeneous plasma with a uniform background magnetic field $\mathbf{B}_0$, small-amplitude fluctuations, where linear terms ($\delta V_{\rm rms}V_{\rm A}$, $\delta B_{\rm rms}B_0$) dominate over nonlinear terms ($\delta V_{\rm rms}^2$, $\delta B_{\rm rms}^2$), can be decomposed into three linear MHD eigenmodes: the incompressible Alfvén mode, and the compressible fast and slow modes \citep{Hollweg1975,Verscharen2019}. Here, $V_{\rm A}$ is the Alfvén speed, magnetic field strength $B_0 = |\mathbf{B}_0|$, and $\delta V_{\rm rms}$ and $\delta B_{\rm rms}$ are the root-mean-square (rms) of velocity and magnetic field fluctuations. For each mode, the temporal power spectrum at a given wavevector $\mathbf{k}$ peaks at its eigenfrequency $\omega_{\rm wave}$. In the linear regime, this peak reduces to a $\delta$-function at $\omega_{\rm wave}$, satisfying respective linear dispersion relations.

In turbulent plasmas, fluctuations on MHD scales are not expected to strictly satisfy linear dispersion relations due to nonlinear interactions. As a result, the temporal power spectrum broadens from an idealized $\delta$-function at $\omega_{\rm wave}$ into a finite-width feature \citep{Zhao2024a,Yuen2025}. Based on the strength of nonlinear coupling, MHD turbulence is classified into weak and strong regimes \citep{Lazarian1999,Meyrand2016,Zhao2024a}. In the weak-turbulence regime, interactions among counter-propagating wave packets are sufficiently weak that fluctuating energy remains predominantly near $\omega_{\rm wave}$ \citep{Galtier2000}. With increasing nonlinear coupling strength, fluctuations gradually attain a critical balance between the linear wave timescale, $\tau_{\rm wave}=1/\omega_{\rm wave}$, and the nonlinear timescale, $\tau_{\rm nl}$, marking a transition to the strong-turbulence regime \citep{Shebalin1983,Goldreich1995}. In this regime, off-dispersion fluctuations are expected to dominate the spectrum \citep{Meyrand2016,Zhao2024a}, accounting for approximately $75\%$–$80\%$ of the total power \citep{Gan2022}. Furthermore, theory predicts mode-dependent asymmetric frequency broadening \citep{Yuen2025}: (i) Alfv\'en and slow fluctuations develop a pronounced low-frequency continuum dominated by sufficiently small $k_\parallel$, while retaining spectral peaks near $\omega_{\rm wave}$ at finite  $k_\parallel$, where $k_\parallel$ denotes wavenumbers parallel to $\mathbf{B}_0$. (ii) In contrast, fast modes remain predominantly wave-like, exhibiting modest frequency broadening near $\omega_{\rm wave}$ and contributing little power to the low-frequency continuum; accordingly, they exhibit comparatively weak nonlinearity across the inertial range \citep{Galtier2003,Yuen2025,Hou2025}.

Beyond mode-dependent nonlinear frequency broadening, cascade behaviors in wavenumber space further shape the spatiotemporal properties of compressible MHD turbulence. In the weak-turbulence regime, Alfvén modes transfer energy to higher $k_\perp$ through resonant three-wave interactions \citep{Galtier2000}, where $k_\perp$ denotes the wavenumber perpendicular to $\mathbf{B}_0$. In the strong-turbulence regime, Alfvén modes satisfy the critical balance ($\tau_{\rm wave}\sim\tau_{\rm nl}$), leading to an anisotropic cascade with $k_\parallel \propto k_\perp^{2/3}$ \citep{Goldreich1995}. A comprehensive theoretical framework for compressible MHD turbulence that self-consistently incorporates slow and fast modes remains incomplete. Nevertheless, slow modes are expected to be passively cascaded by Alfvénic turbulence with similar anisotropy \citep{Lithwick2001,Cho2003,Schekochihin2009,Makwana2020}. In contrast, fast modes remain nearly isotropic, exhibiting an energy spectrum $E(k)\propto k^{-3/2}$ and being modulated by collisionless transit-time damping that strengthens with obliquity at small scales \citep{Yan2004,Petrosian2006,Suzuki2006}, where $k=|\mathbf{k}|$. Despite theoretical and numerical advances, direct observational characterizations of mode-resolved spatiotemporal properties across weak and strong turbulence regimes remain limited.

Reliable mode decomposition is a prerequisite for investigating the mode-resolved spatiotemporal properties of compressible MHD turbulence. \citet{Glassmeier1995} decomposed the time series of fluctuations into three sets of forward and backward propagating MHD eigenmodes in frequency space by projecting the fluctuating vector to eigenvectors. \citet{Cho2002,Cho2003} performed mode-decomposition in wavenumber space by projecting turbulent fluctuations onto the displacement vectors of three MHD eigenmodes. \citet{Zank2023} additionally incorporated nonpropagating entropy and magnetic-island modes. However, these approaches rely on linear MHD dispersion relations, whereas nonlinear interactions and the associated frequency broadening drive turbulent fluctuations away from linearity \citep{Zhao2024a,Yuen2025}. Such intrinsic departures may significantly limit the applicability of linear-based decomposition methods, particularly in the strong-turbulence regime. To overcome this limitation, \citet{Zhao2026} recently introduced a polarization-based mode-decomposition method that separates Alfvénic and compressible fluctuations without enforcing linear MHD dispersion relations, but it does not distinguish between fast and slow modes.

In this study, to separate fast and slow modes, we further advance the polarization-based mode-decomposition method of \citet{Zhao2026}, augmented by incorporating the phase correlation between fluctuations in magnetic field strength $\delta |\mathbf{\tilde{B}}|$ and plasma density $\delta \tilde{N}$. Using joint observations from the four Cluster spacecraft in Earth’s magnetosheath and the advanced polarization-based approach, we obtain four-dimensional spatiotemporal power spectra of compressible MHD turbulence, including Alfvén, slow, and fast modes, without invoking assumptions such as the Taylor hypothesis \citep{Taylor1938}. We further provide the first quantitative assessment of how frequency broadening depends on nonlinear interactions, enabling a direct test of the weak-to-strong transition in compressible turbulence.

\section{Mode-Decomposition Method}

Magnetic field data are obtained from the Fluxgate Magnetometer (FGM) \citep{Balogh1997} at $22.5$ Hz, proton plasma parameters from the Cluster Ion Spectrometry’s Hot Ion Analyzer (CIS-HIA) \citep{Reme2001}, and electron density from the Plasma Electron And Current Experiment (PEACE) \citep{Johnstone1997} and the Waves of High frequency and Sounder for Probing of Electron density by Relaxation (WHISPER) \citep{Decreau1997}.

In turbulent plasma, MHD fluctuations consist of incompressible Alfvén modes, compressible fast and slow (magnetosonic) modes, as well as fluctuations that deviate from the linear dispersion relations owing to nonlinear interactions \citep{Zhao2024a,Yuen2025}. The linear dispersion relations are given by $f_{\rm A}=k_\parallel V_{A}/(2\pi)$, $f_{\rm fast}=kV_{\rm ph,+}/(2\pi)$, and $f_{\rm slow}=kV_{\rm ph,-}/(2\pi)$, where $f_{\rm A}$, $f_{\rm fast}$, and $f_{\rm slow}$ denote the eigenfrequencies of Alfvén, fast, and slow modes, respectively \citep{Hollweg1975}. The phase speeds of fast and slow modes are given by 
\begin{eqnarray}
V_{\rm ph,\pm}^2=\frac{1}{2}\{(V_{\rm S}^2 + V_{\rm A}^2) \pm [(V_{\rm S}^2 + V_{\rm A}^2)^2 - 4V_{\rm S}^2V_{\rm A}^2\rm cos^2\theta]^\frac{1}{2}\},
\end{eqnarray} 
where `$+$' and `$-$' signs correspond to the fast and slow modes, respectively; $V_{\rm S}$ is the sound speed, and $\theta$ is the angle between $\mathbf{k}$ and $\mathbf{B}_0$.

We decompose small-amplitude fluctuations into Alfvénic and compressible modes based on their polarization properties \citep{Zhao2026}. For compressible fluctuations, oblique fast (slow) modes exhibit in-phase (anti-phase) correlations between $\delta |\mathbf{\tilde{B}}|$ and $\delta \tilde{N}$ \citep{Hollweg1975,Verscharen2019}. In addition, fast modes remain predominantly wave-like with weak frequency broadening across the inertial range, whereas slow modes increasingly deviate from their linear dispersion relations as nonlinearity increases \citep{Yuen2025}. Motivated by these distinctions, we isolate fast modes by integrating power within a $\pm30\%$ frequency band centered at $f_{\rm fast}$, taking into account the expected in-phase correlation. The remaining compressible fluctuations are therefore likely dominated by slow modes, characterized by anti-phase correlations, along with residual in-phase fluctuations that deviate from the fast-mode dispersion relation.

\begin{figure}[t!]
\centering
\includegraphics[scale=0.5]{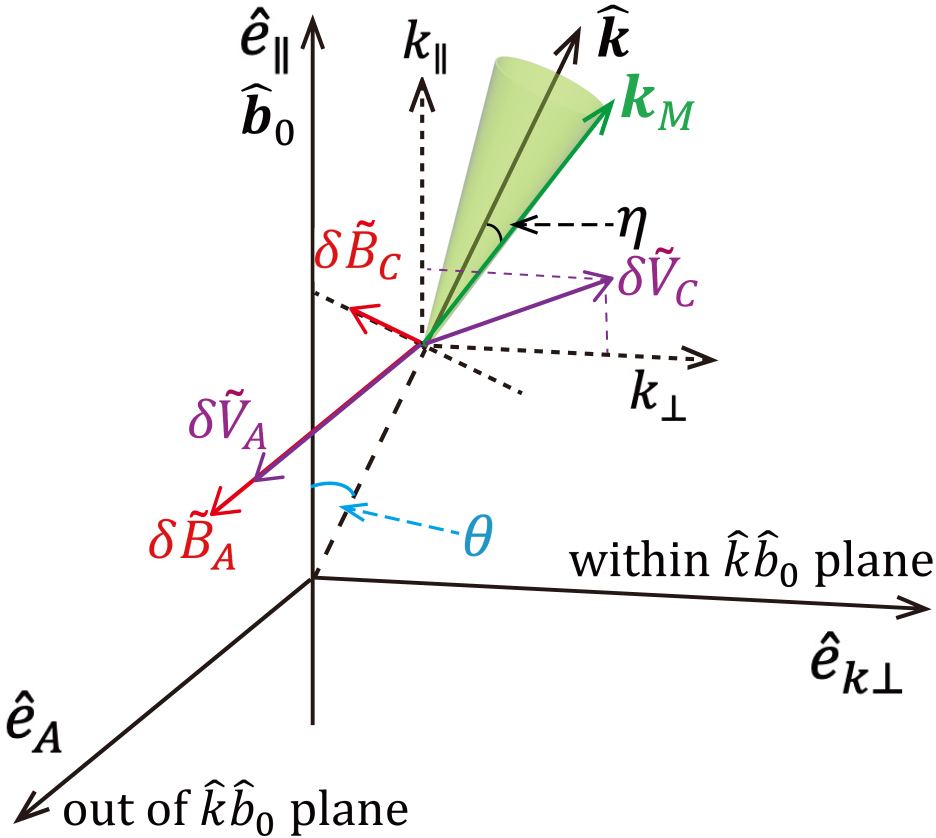}
 \caption{\label{fig:coordinate} Fluctuations in the $\hat{k}\hat{b}_0$ coordinates. The orthonormal basis vectors are $\hat{\mathbf{e}}_{\parallel}=\hat{\mathbf{b}}_0$, $\hat{\mathbf{e}}_{k\perp} = \hat{\mathbf{b}}_0\times(\hat{\mathbf{k}}\times\hat{\mathbf{b}}_0) / |\hat{\mathbf{b}}_0\times(\hat{\mathbf{k}}\times\hat{\mathbf{b}}_0)|$, and $\hat{\mathbf{e}}_{A} = \hat{\mathbf{e}}_{k\perp}\times\hat{\mathbf{e}}_{\parallel}$. Alfvénic magnetic and velocity fluctuations ($\delta \tilde{B}_A$ and $\delta \tilde{V}_{\rm A}$) are polarized perpendicular to the $\hat{k}\hat{b}_0$ plane, whereas compressible fluctuations ($\delta\tilde{B}_C$ and $\delta \tilde{V}_C$) are polarized within the $\hat{k}\hat{b}_0$ plane. $\theta$ is the angle between $\hat{\mathbf{k}}$ and $\hat{\mathbf{b}}_0$, and $\eta$ is the angle between timing-derived ${\mathbf{k}}_M$ and SVD-derived $\hat{\mathbf{k}}$.}
 \end{figure}

The applicability conditions of this polarization-based decomposition technique are as follows: (1) the small-amplitude assumption must hold, i.e., the linear terms dominate over the nonlinear terms, such that the fluctuations can be decomposed into the incompressible Alfvén mode and the compressible fast and slow modes; (2) one mode must be dominant so that the local $\hat{k}\hat{b}_0$ coordinate system can be constructed more reliably; (3) the analysis interval should remain temporally stable and spatially homogeneous, with well-developed fluctuations.

We analyze small-amplitude fluctuations in the $\hat{k}\hat{b}_0$ coordinates defined by the unit wavevector $\hat{\mathbf{k}}=\frac{\mathbf{k}}{|\mathbf{k}|}$ and unit background magnetic field $\hat{\mathbf{b}}_0=\mathbf{B}_0/|\mathbf{B}_0|$ in Figure~\ref{fig:coordinate} \citep{Cho2003,Zhao2024b,Zhao2024a}. The unit wavevector $\hat{\mathbf{k}}$ is determined using singular value decomposition (SVD) based on the linearized Gauss’s law for magnetism \citep{Santolik2003}. Because turbulent eddies interact with the local rather than the global magnetic field \citep{Cho2000,Zhao2022,Yuen2023}, we define the local background magnetic field as $\mathbf{B}_0(t,\tau)=(\mathbf{B}(t-\tau/2)+\mathbf{B}(t+\tau/2))/2$, where $t$ is time, $\tau=1/f_{\rm sc}$ is the timescale, and $f_{\rm sc}$ is the spacecraft-frame frequency. The procedure for constructing the spatiotemporal power spectra is described below.

First, the time series of fluctuations are transformed using the Morlet-wavelet transform \citep{Grinsted2004}, yielding Fourier-space fluctuations in geocentric-solar-ecliptic (GSE) coordinates: $\mathbf{\delta \tilde{B}}(t,f_{\rm sc})=[\delta \tilde{B}_{\rm X},\delta \tilde{B}_{\rm Y},\delta \tilde{B}_{\rm Z}]$, $\mathbf{\delta \tilde{V}}_{\rm p}(t,f_{\rm sc})=[\delta \tilde{V}_{\rm pX},\delta \tilde{V}_{\rm pY},\delta \tilde{V}_{\rm pZ}]$, $\delta \tilde{N}_{\rm p}(t,f_{\rm sc})$, and $\delta \tilde{N}_{\rm e}(t,f_{\rm sc})$, where the tilde denotes the wavelet coefficient of the fluctuations. The whole time interval is segmented into overlapping windows of $t_{\mathrm{win}} = 5$ hours with a 30-minute shift to ensure adequate statistical sampling, where $t_{\mathrm{win}}$ is the window duration. All magnetic field and plasma parameters are interpolated to a uniform resolution of 1 sample/s. To reduce edge effects from finite time series, each wavelet transform is performed over a $2t_{\mathrm{win}}$ interval, with only the central $t_{\rm win}$ retained for analysis.

Second, based on MHD theory, Alfvénic velocity fluctuations $\delta \mathbf{\tilde{V}}_{\rm A}$ are identified by their incompressibility, $\hat{\mathbf{k}}\cdot\delta\mathbf{\tilde{V}}_{\rm A}=0$, and perpendicularity to the local mean magnetic field, $\mathbf{B}_0\cdot\delta\mathbf{\tilde{V}}_{\rm A}=0$. In contrast, compressible velocity fluctuations $\delta \mathbf{\tilde{V}}_{\rm C}$ lie in the $\hat{k}\hat{b}_0$ plane and include contributions from fast and slow modes. 
Based on the linearized induction equation, Alfv\'enic magnetic fluctuations $\delta \mathbf{\tilde{B}}_{\rm A}$ are polarized along $\hat{\mathbf{e}}_{\rm A}$ and are given by $\delta \mathbf{\tilde{B}}_{\rm A}= \mathbf{\delta \tilde{B}} \cdot \hat{\mathbf{e}}_{\rm A}$, whereas compressible magnetic fluctuations $\delta \mathbf{\tilde{B}}_{\rm C}$ lie within the $\hat{k}\hat{b}_0$ plane and are polarized along $\hat{\mathbf{e}}_{\rm C}=\hat{\mathbf{e}}_{\rm A}\times\hat{\mathbf{k}}$, i.e., $\delta \mathbf{\tilde{B}}_{\rm C}= \mathbf{\delta \tilde{B}} \cdot \hat{\mathbf{e}}_{\rm C}$ (Figure~\ref{fig:coordinate}). Proton and electron density fluctuations ($\delta \tilde{N}_{\rm p}$ and $\delta \tilde{N}_{\rm e}$) arise entirely from compressible modes. The associated velocity, magnetic, and density powers are calculated as 

\begin{align}
P_{\rm KA}(t,f_{\rm sc}) &= \delta \tilde{V}_{\rm A}\delta \tilde{V}_{\rm A}^*,\\
P_{\rm KC}(t,f_{\rm sc}) &= \delta \tilde{V}_{\rm C}\delta \tilde{V}_{\rm C}^*
= \delta \tilde{V}_{\parallel}\delta \tilde{V}_{\parallel}^* + \delta \tilde{V}_{k\perp}\delta \tilde{V}_{k\perp}^*,\\
P_{\rm BA}(t,f_{\rm sc}) &= \delta \tilde{B}_{\rm A}\delta \tilde{B}_{\rm A}^*,\\
P_{\rm BC}(t,f_{\rm sc}) &= \delta \tilde{B}_{\rm C}\delta \tilde{B}_{\rm C}^*,\\
P_{\rm N_p}(t,f_{\rm sc})&= \delta \tilde{N}_{\rm p}\delta \tilde{N}_{\rm p}^*,\\
P_{\rm N_e}(t,f_{\rm sc})&= \delta \tilde{N}_{\rm e}\delta \tilde{N}_{\rm e}^*,
\end{align}
where $\delta \tilde{V}_{\parallel}$ and $\delta \tilde{V}_{k\perp}$ are proton velocity fluctuations along $\hat{\mathbf{e}}_{\parallel}$ and $\hat{\mathbf{e}}_{k\perp}$, respectively.

Third, we determine wavevectors using multi-spacecraft timing analysis \citep{Pincon2008} for intervals with a tetrahedron quality factor $\rm TQF>0.8$, ensuring reliable and accurate wavenumber estimation (see \cite{Zhao2024b} for details). Unlike singular value decomposition (SVD) technique, which provides only a wavevector direction $\mathbf{\hat{k}}$ corresponding to the normal of the composite $\delta\mathbf{\tilde{B}}$ plane formed by superposed modes, the timing analysis yields the full wavevector $\mathbf{k}_{\rm M}$ from the inter-spacecraft phase differences of $\delta \mathbf{\tilde{B}}_{\rm M}$, where $\rm M=A$ and $\rm M=C$ denote Alfvénic and compressible fluctuations, respectively. Because modes with distinct dispersion relations and wavevectors can coexist at the same frequency, $\mathbf{k}_{\rm M}$ ($\mathbf{k}_{\rm A}$ or $\mathbf{k}_{\rm C}$) is not necessarily aligned with the SVD-derived $\mathbf{\hat{k}}$. We therefore regard the timing analysis results as reliable only when $\mathbf{k}_{\rm M}$ is closely aligned with $\mathbf{\hat{k}}$, and restrict our analysis to cases where the angle $\eta$ between them is below a prescribed threshold (Figure~\ref{fig:coordinate}).

Fourth, the power spectra are transformed into the rest frame of plasma flow by correcting the Doppler shift, $f_{\mathrm{rest}}=f_{\rm sc}-\mathbf{k_M}\cdot\mathbf{V}_{\rm p}/(2\pi)$, where $f_{\mathrm{rest}}$ is the frequency in the rest frame of plasma flow, $\mathbf{V}_{\rm p}$ is the proton bulk velocity, and the spacecraft velocity is considered to be negligible. We adopt an absolute-frequency representation, 
\begin{align}
   (f_{\mathrm{rest}}, \mathbf{k}_{\rm M}) =
   \begin{cases}
       (f_{\mathrm{rest}}, \mathbf{k}_{\rm M}) & \text{if  } f_{\mathrm{rest}} > 0, \\
       (-f_{\mathrm{rest}}, -\mathbf{k}_{\rm M}) & \text{if  } f_{\mathrm{rest}} < 0,
   \end{cases}
\end{align}
which ensures positive frequencies by simultaneously reversing the signs of both frequency and wavevector when $f_{\mathrm{rest}}<0$.

Finally, we construct a set of $N_f\times N_{k_\parallel}\times N_{k_\perp}$ bins to obtain spatiotemporal power spectra of velocity, magnetic, and density fluctuations in the plasma flow frame. Each bin subtends approximately constant $f_{\mathrm{rest}}$, $k_\parallel$, and $k_\perp$, with widths $df_{\mathrm{rest}}$, $dk_\parallel$, and $dk_\perp$. The wavenumbers are calculated as $k = |\mathbf{k}_M|$, $k_\parallel=\mathbf{k}_{M}\cdot\mathbf{\hat{b}}_0$, and $k_\perp=\sqrt{\mathbf{k}_M^2-k_\parallel^2}$. To focus on inertia-range MHD fluctuations, we restrict the analysis to $4/t_{\rm{win}}<f_{\mathrm{rest}}<f_{\rm cp}/2$ Hz and $1/(100d_{\rm sc})<k<{\rm \min}(1/\rm \max(d_p,r_{\rm cp}),\pi/d_{\rm sc})$. Here, $f_{\rm cp}$ is proton gyrofrequency, $d_{\rm sc}$ is the spacecraft separation, $d_{\rm p}$ is the proton inertial length, $r_{\rm cp}$ is the proton gyroradius, and $\max$ and $\min$ denote the maximum and minimum. These bounds ensure adequate averaging over multiple wave periods and account for the intrinsic limitation of multi-spacecraft techniques, which are sensitive to spatial scales comparable to the spacecraft separation \citep{Pincon2008}.

\section{Results}

During 23:00-10:00 UT on 2–3 December 2003, four Cluster spacecraft traversed the flank of Earth’s magnetosheath in a near-ideal tetrahedral configuration with a tetrahedron quality factor $\rm TFQ=0.97$ and spacecraft separation $d_{\rm sc}=200$ km. During this interval, the magnetic field and plasma parameters remain temporally stable and spatially homogeneous, with well-developed fluctuations, providing an ideal plasma environment for investigating the spatiotemporal properties of compressible MHD turbulence (see an overview in Figure~\ref{fig:overview} in Appendix~\ref{Appendix:overview}). This interval has previously been analyzed to investigate the weak-to-strong transition in Alfvénic turbulence \citep{Zhao2024a}. Building upon that work, we revisit the same interval to extend the analysis to compressible MHD turbulence and, importantly, to quantify nonlinear frequency broadening, enabling a direct and consistent comparison between Alfvénic and compressible modes.

\subsection{Spatiotemporal power spectra of magnetic fluctuations}

 \begin{figure}[t!]
\centering
\includegraphics[scale=0.15]{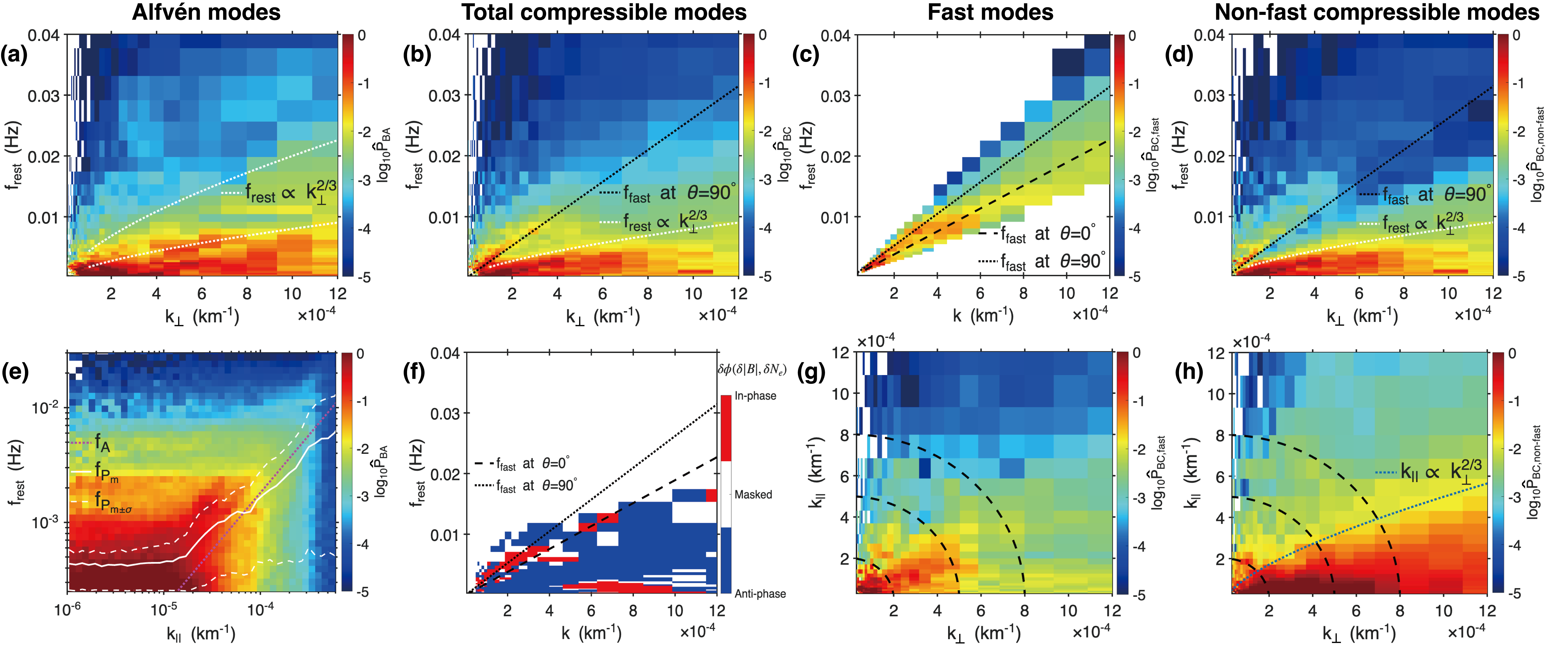}
 \caption{\label{fig:2D_spec} Spatiotemporal power spectra of magnetic fluctuations in the plasma flow frame. (a,b,d) $f_{\mathrm{rest}}-k_\perp$ distributions of Alfvénic power ($\hat{P}_{\rm BA}$), total compressible power ($\hat{P}_{\rm BC}$), and non-fast compressible power ($\hat{P}_{\rm BC,non\text{-}fast}$). White dotted curves denote the scaling $f_{\rm rest} \propto k_\perp^{2/3}$. Black dashed (dotted) lines denote the fast-mode dispersion relations at $\theta=0^\circ$ ($90^\circ$), respectively. (c) $f_{\mathrm{rest}}-k$ distribution of fast-mode power ($\hat{P}_{\rm BC,fast}$). (e) $f_{\mathrm{rest}}-k_\parallel$ distribution of $\hat{P}_{BA}$. Pink dotted line denotes the Alfvén-mode dispersion relation. White solid and dashed curves indicate the frequency of the mean power spectrum ($f_{\rm P_m}$) and $\pm\sigma$ confidence intervals ($f_{\rm P_{m\pm\sigma}}$), where $\sigma$ is the standard deviation of the power. (f) Phase correlation between fluctuating magnetic field strength ($\delta |\mathbf{\tilde{B}}|$) and PEACE electron density ($\delta \tilde{N}_e$), measured by Cluster-2. In-phase (anti-phase) regions correspond to phase differences within $\pm80^\circ$ around $0^\circ$ ($180^\circ$), leaving a narrow transition region. (g,h) $k_\parallel-k_\perp$ distributions of $\hat{P}_{\rm BC,fast}$ and $\hat{P}_{\rm BC,non\text{-}fast}$. Black dashed curves denote isotropic contours at $k=2\times10^{-4}$, $5\times10^{-4}$, and $8\times10^{-4}$ $\rm km^{-1}$. Blue dotted curve denotes the scaling $k_{\rm \perp} \propto k_\perp^{2/3}$. Panels with the same format are normalized by the same factor, and values smaller than $10^{-6}$ are set to NaN. The data are binned into $N_f\times N_{k_\parallel}\times N_{k_\perp}=50\times50\times50$ with $\eta\leq10^\circ$, where $\eta$ is the angle between wavevectors determined by SVD and timing analysis (Figure~\ref{fig:coordinate}).}
 \end{figure}

Benefiting from multi-spacecraft observations, the frequency and wavenumber spectra are measured independently rather than inferred, enabling us to directly investigate the spatiotemporal distribution of energy across different wave modes. Figure~\ref{fig:2D_spec} shows the spatiotemporal power spectra of the magnetic field in the plasma flow frame, which are the most robust owing to their higher temporal resolution. Corresponding power spectra for velocity, density, and additional magnetic components, exhibiting similar characteristics, are shown in Figure~\ref{fig:v_n_spec} in Appendix~\ref{Appen_B}.

\subsubsection{Alfvén modes}

Figure~\ref{fig:2D_spec}(a) shows that $f_{\rm rest}$-$k_\perp$ distributions of Alfvénic magnetic power, $P_{\rm BA}(f_{\mathrm{rest}},k_\perp)=\int P_{\rm BA}(f_{\mathrm{rest}},k_\parallel,k_\perp)dk_\parallel$, agrees well with the scaling $f_{\mathrm{rest}}\propto k_\perp^{2/3}$. Since $f_{\mathrm{rest}}\propto k_\parallel$ for Alfvén modes, this directly implies the anisotropic scaling $k_\parallel\propto k_\perp^{2/3}$. This relation is consistent with the critical-balance scaling for strong Alfvénic turbulence \citep{Goldreich1995} and is independently confirmed in Figure 2 of \citet{Zhao2024a}. However, that study did not resolve distinct $k_\parallel$ regimes; here, we extend the analysis by explicitly distinguishing them.

Unexpectedly, at large parallel scales, quasi-two-dimensional (quasi-2D) structures are observed that fall outside the theoretical frameworks of both strong and weak turbulence. As shown in Figure~\ref{fig:2D_spec}(e), for $k_\parallel<10^{-5}$ $\rm km^{-1}$, the $f_{\rm rest}$-$k_\parallel$ distribution of Alfvénic magnetic power, $P_{\rm BA}(f_{\mathrm{rest}},k_\parallel)=\int P_{\rm BA}(f_{\mathrm{rest}},k_\parallel,k_\perp)dk_\perp$, exhibits a nearly constant ultra-low-frequency component at $f_{\rm rest}\sim5\times10^{-4}$ Hz. This feature is indicated by the nearly constant power-weighted mean frequency $f_{\rm P_m}$ and its $\pm \sigma$ bounds (white curves). The weak dependence of $f_{\rm P_m}$ on $k_\parallel$, together with the decrease of $P_{\rm BA}$ with increasing $k_\perp$ (Figure~\ref{fig:1D_Alfven}(a)), indicates that the dynamics of these Alfvénic fluctuations are governed by perpendicular scales, demonstrating that these fluctuations with $k_\parallel\ll k_\perp$ correspond to quasi-2D magnetic structures. The compressible magnetic power exhibits similar behavior (Figure~\ref{fig:v_n_spec}(g)), indicating that compressible fluctuations also contribute significantly to low-frequency, large-scale quasi-2D magnetic structures. The ultra-low-frequency behavior is consistent with the strongly nonlinear regime ($\chi=\frac{\tau_{\rm wave}}{\tau_{\rm nl}}\gg1$), where $\omega_{\rm wave}\sim0$, and is observed for both Alfvén and slow modes in simulations \citep{Yuen2025}. Based on an examination of time-series data in the magnetosheath, we identified mirror modes as the dominant structures from their anticorrelation between magnetic and thermal pressures, whereas some events were classified as discontinuities associated with the interplanetary magnetic field (IMF). The observational signatures of coherent structures and turbulent ultra-low-frequency continua can overlap in frequency–wavenumber space. Therefore, we cannot exclude the possibility that the observed low-frequency continuum includes contributions from coherent structures, such as pressure-balanced structures and mirror modes \citep{Balikhin2010,Zank2023}, in addition to strongly nonlinear turbulence and quasi-2D dynamics. The physical nature of these ultra-low-frequency fluctuations is beyond the scope of this work and will be investigated in future studies.

For $k_\parallel>10^{-5}$ $\rm km^{-1}$, $f_{\rm P_m}$ of Alfvénic magnetic power remains close to $f_A$ (pink dotted line in Figure~\ref{fig:2D_spec}(e)), indicating frequency broadening centered at $f_A$. In contrast, $f_{\rm P_m}$ of compressible magnetic power deviates significantly from $f_{\rm slow}$ and is preferentially shifted to lower frequencies, particularly at $k_\parallel>7\times10^{-5}$ $\rm km^{-1}$ (Figure~\ref{fig:v_n_spec}(f-h) in Appendix~\ref{Appen_B}). This behavior likely reflects greater nonlinearity in slow modes, consistent with simulations that show enhanced low-frequency power in these modes \citep{Yuen2025}. 

\subsubsection{Compressible modes}

Figure~\ref{fig:2D_spec}(b) shows that compressible magnetic power, $P_{\rm BC}(f_{\mathrm{rest}},k_\perp)=\int P_{\rm BC}(f_{\mathrm{rest}},k_\parallel,k_\perp)dk_\parallel$, exhibits two distinct components. (1) One follows the fast-mode dispersion relations (black dotted line). (2) The other is concentrated at low frequencies and agrees with the scaling $f_{\mathrm{rest}}\propto k_\perp^{2/3}$, similar to that found in $P_{\rm BA}$ (Figure~\ref{fig:2D_spec}(a)).

As a preliminary diagnostic of compressible mode composition, we examine the phase correlation between $\delta |\mathbf{\tilde{B}}|$ and PEACE electron density $\delta \tilde{N}_{\rm e}$. Figure~\ref{fig:2D_spec}(f) shows that anti-phase fluctuations dominate most of the spatiotemporal domain, consistent with the anti-phase correlation of slow modes. By contrast, in-phase fluctuations cluster in two regions: near the fast-mode dispersion relations (black lines) and in a low-frequency range at $k=[5\times10^{-4},10^{-3}]$ $\rm km^{-1}$, where inferred phase speeds $V_{\rm ph}=[23,30]$ $\rm kms^{-1}$ ($V_{\rm ph}/V_{\rm A}\sim0.2$) suggest slowly propagating magnetic structures in Earth's magnetosheath. These structures contribute to compressible power below $f_{\rm slow}$ at $k>10^{-4}$ $\rm km^{-1}$ (Figures~\ref{fig:v_n_spec}(f-h)). The phase-correlation method is described in Figure~\ref{fig:phase_BN} and Appendix~\ref{Appen_C}, where CIS–HIA and WHISPER density results show similar behavior.

Guided by these in-phase-correlation signatures, we extract fast-mode fluctuations as $P_{\rm BC,fast}(f_{\mathrm{rest}},k) = P_{\rm BC}(0.7f_{\rm fast}< f_{\mathrm{rest}}< 1.3f_{\rm fast},k)$. The influence of fast-mode selection criteria on the results is discussed in Appendix~\ref{Appen_D}. Figure~\ref{fig:2D_spec}(c) shows that $P_{\rm BC,fast}$ is primarily distributed at $k<8\times10^{-4}$ $\rm km^{-1}$. At larger $k$, $P_{\rm BC,fast}$ becomes progressively confined to the fast-mode dispersion relation at quasi-parallel propagation, consistent with enhanced collisionless transit-time damping of highly oblique fast modes \citep{Yan2004,Petrosian2006}. Regions of enhanced fast-mode power coincide with in-phase magnetic–density correlations (Figures~\ref{fig:2D_spec}(c,f)), supporting the robustness of fast-mode identification. Figure~\ref{fig:2D_spec}(g) shows the fast-mode magnetic power, $P_{\rm BC,fast}(k_\parallel,k_\perp) = \int_{0.7f_{\rm fast}}^{1.3f_{\rm fast}} P_{\rm BC}(f_{\mathrm{rest}},k_\parallel,k_\perp)df_{\mathrm{rest}}$, which exhibits weak dependence on propagation angle $\theta$ for $k<8\times10^{-4}$ $\rm km^{-1}$, consistent with the isotropic cascade of fast modes in the absence of damping \citep{Cho2002,Hou2025}. For $k\geq8\times10^{-4}$~$\rm km^{-1}$, $P_{\rm BC,fast}$ becomes more anisotropic, likely due to contamination from anti-phase magnetic-density fluctuations. As phase correlations become unstable at larger $k$ (white regions in Figure~\ref{fig:2D_spec}(f)), fast-mode fluctuations cannot be reliably identified at these scales and are excluded from further analysis.

The remaining compressible magnetic fluctuations are classified as non-fast modes, defined by $P_{\rm BC,non\text{-}fast}=P_{\rm BC}-P_{\rm BC,fast}$, and exhibit predominantly anti-phase magnetic–density correlations (Figure~\ref{fig:2D_spec}(f)), consistent with slow and magnetic-mirror modes. As shown in Figure~\ref{fig:2D_spec}(d), $P_{\rm BC,non\text{-}fast}$ is concentrated at ultra-low frequencies ($f_{\mathrm{rest}}<5\times10^{-3}$ Hz) and is preferentially elongated toward higher $k_\perp$ than the fast-mode component, consistent with simulations \citep{Gan2022,Yuen2025}. It also follows the anisotropic scaling $k_\parallel \propto k_\perp^{2/3}$ (Figure~\ref{fig:2D_spec}(h)), characteristic of the scale-dependent cascade of slow modes \citep{Cho2002,Makwana2020}. Together, these results indicate that the non-fast compressible fluctuations are predominantly associated with slow modes.

We note that this procedure does not completely remove fast-mode contributions. This is partly because a fraction of in-phase magnetic-density fluctuations at ultra-low frequencies that do not follow fast-mode dispersion relations are included in the non-fast component (Figures~\ref{fig:2D_spec}(d,f)), and partly because fast-mode frequency broadening is scale-dependent (Figure~\ref{fig:1D_comp}). Nevertheless, spatiotemporal characteristics of $P_{\rm BC,fast}$ and $P_{\rm BC,non\text{-}fast}$ remain consistent with theoretical expectations for fast and slow modes, respectively, indicating that this approach provides an effective, though not exact, separation and captures the essential dynamics.

\subsection{Nonlinear frequency broadening of three MHD eigenmodes}\label{nonlinear}

\begin{figure}[t!]
\centering
\includegraphics[scale=0.2]{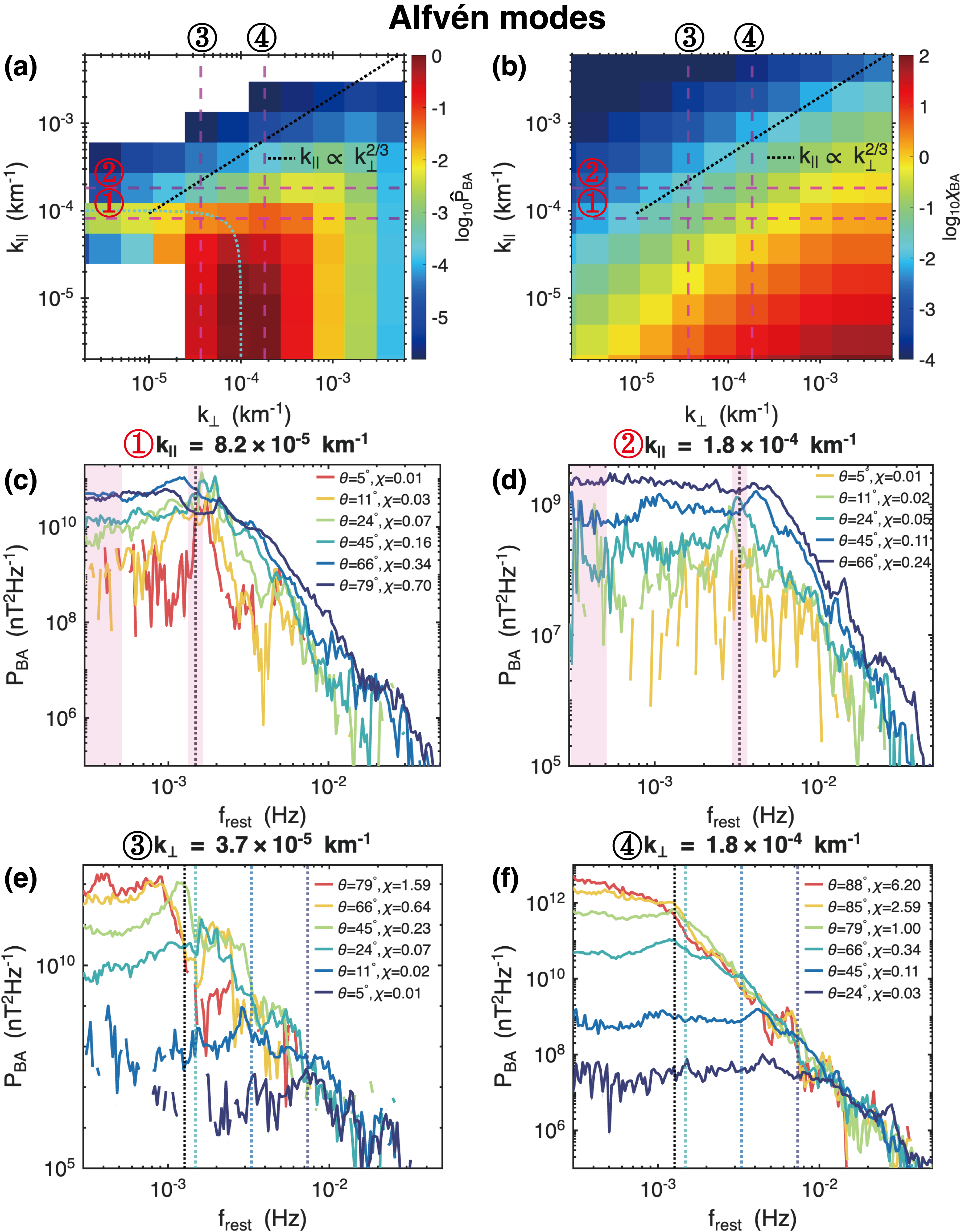}
 \caption{\label{fig:1D_Alfven} Alfvénic magnetic power in the plasma flow frame. (a) $k_\parallel-k_\perp$ distribution of $\hat{P}_{\rm BA}$, where $\hat{P}_{\rm BA}=P_{\rm BA}/P_{\rm BA,max}$ is normalized by the maximum over all ($k_\parallel,k_\perp$) bins. Blue contour marks isotropy at $k=10^{-4}$ $\rm km^{-1}$. Black dotted lines denote the scaling $k_{\rm \parallel} \propto k_\perp^{2/3}$. (b) $k_\parallel-k_\perp$ distribution of the Alfvénic nonlinearity parameter $\chi_{\rm BA}$. (c,d) $P_{\rm BA}$ at fixed $k_\parallel$, with color indicating $k_\perp$. The angle $\theta= \rm arctan(k_\perp/k_\parallel)$. Dotted lines indicate $f_A$ calculated with $k_\parallel=8.2\times10^{-5}$ and $1.8\times10^{-4}$ $\rm km^{-1}$. Pink-shaded regions indicate the quasi-zero-frequency power $P_{\rm BA}(\mathbf{k},f_{\mathrm{rest}}\sim f_{\mathrm{min}})=\frac{\int_{f_{\mathrm{min}}}^{5\times10^{-4} \rm Hz}P_{\rm BA}(\mathbf{k},f_{\mathrm{rest}})df_{\mathrm{rest}}}{5\times10^{-4} \rm Hz-f_{\mathrm{min}}}$ and Alfvén-wave-like power $P_{\rm BA}(\mathbf{k},f_{\mathrm{rest}}\sim f_A)= \frac{\int_{0.9f_{\mathrm{A}}}^{1.1f_{\mathrm{A}}}P_{\rm BA}(\mathbf{k},f_{\mathrm{rest}})df_{\mathrm{rest}}}{0.2f_{\mathrm{A}}}$. (e,f) $P_{\rm BA}$ at fixed $k_\perp$, with color indicating $k_\parallel$. Colored dotted lines show the corresponding $f_A$, and black dotted lines indicate $f_{\rm A0}$ from $k_{\parallel,0}\sim7\times10^{-5}$ $\rm km^{-1}$. The data are binned into $N_f\times N_{k_\parallel}\times N_{k_\perp}=200\times12\times12$ with $\eta\leq30^\circ$ to improve statistical robustness. }
 \end{figure}

\subsubsection{Alfvén modes}

Figure~\ref{fig:1D_Alfven}(b) shows that the Alfvénic nonlinearity parameter, $\chi_{\rm BA}=k_\perp\delta \tilde{B}_A/(k_\parallel B_0)$, increases with increasing $k_\perp$ or decreasing $k_\parallel$, indicating stronger nonlinear interactions. The Alfvénic magnetic energy density is calculated as $\delta B_A^2(k_\parallel,k_\perp)=\int_0^\infty \int_{k_\parallel}^\infty\int_{k_\perp}^\infty P_{BA}(f_{\mathrm{rest}},k_\parallel,k_\perp)df_{\mathrm{rest}}dk_\parallel dk_\perp$. Figures~\ref{fig:1D_Alfven}(c-f) show Alfvénic magnetic power spectra $P_{\rm BA}$ at selected ($k_\parallel$, $k_\perp$). For $\chi_{\rm BA}\ll1$, $P_{\rm BA}$ exhibits pronounced peaks near $f_A$, indicative of weakly nonlinear, wave-like behavior. As $\chi_{\rm BA}$ increases, frequency broadening becomes significant: $P_{\rm BA}$ evolves from narrow wave-like peaks to frequency-broadened distributions characterized by a plateau below $f_A$ and a power-law decay at higher frequencies. This evolution is consistent with the weak-to-strong transition in Alfvénic turbulence \citep{Zhao2024a}.

\begin{figure}[t!]
\centering
\includegraphics[scale=0.3]{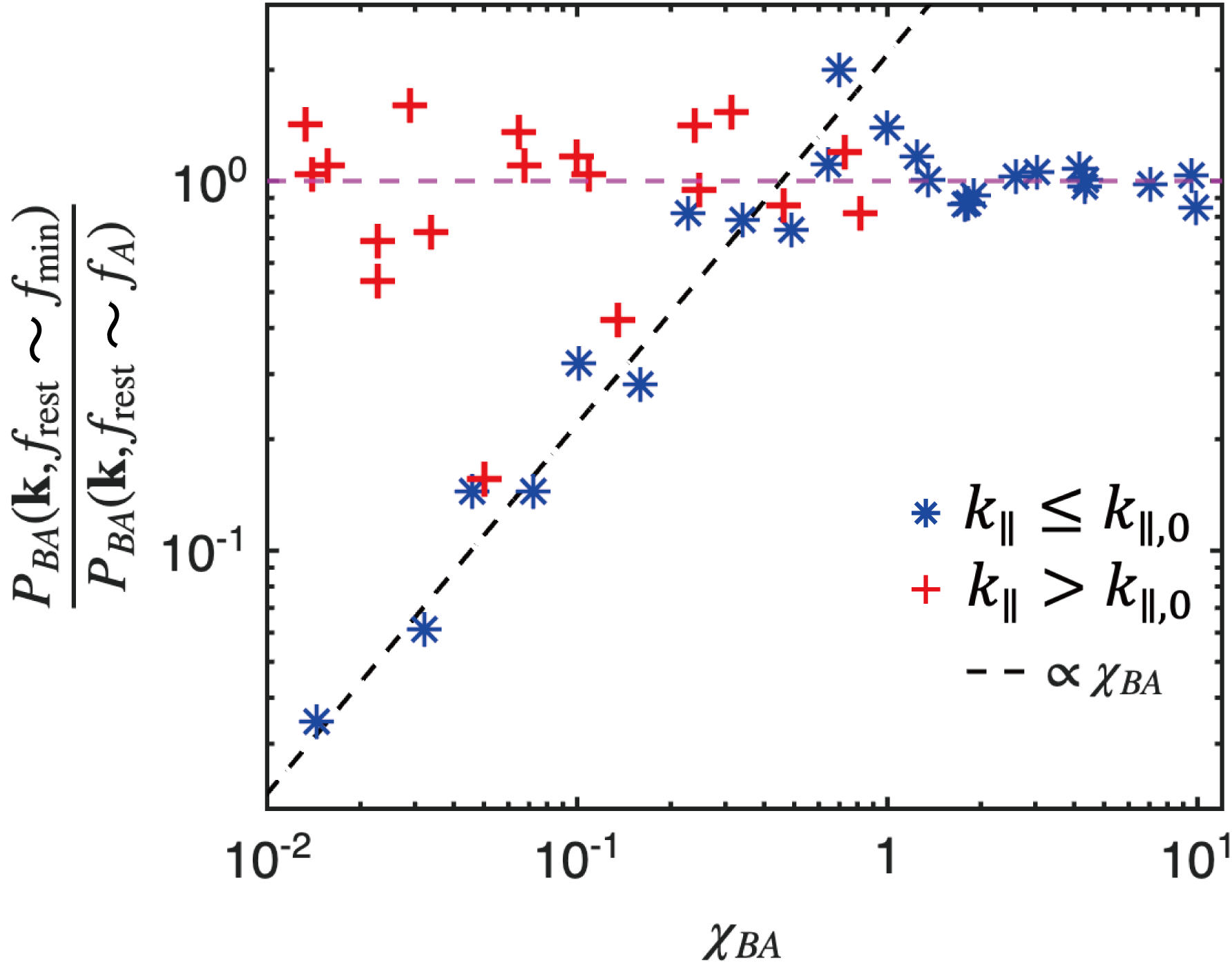}
 \caption{\label{fig:Relationship} Dependence of $\frac{P_{\rm BA}(\mathbf{k},f_{\mathrm{rest}}\sim f_{\mathrm{min}})}{P_{\rm BA}(\mathbf{k},f_{\mathrm{rest}}\sim f_A)}$ on the Alfvénic nonlinearity parameter $\chi_{BA}$. The quasi-zero-frequency power $P_{\rm BA}(\mathbf{k},f_{\mathrm{rest}}\sim f_{\mathrm{min}})$ and Alfvén-wave-like power $P_{\rm BA}(\mathbf{k},f_{\mathrm{rest}}\sim f_A)$ are indicated by the pink-shaded regions in Figures~\ref{fig:1D_Alfven}(c,d). The minimum frequency is $f_{\rm min}=4/t_{\rm win}$. Points with $k_\parallel \leq k_{\parallel,0}$ ($k_\parallel > k_{\parallel,0}$) are marked by asterisks (pluses), respectively. Pink horizontal line indicates $\frac{P_{\rm BA}(\mathbf{k},f_{\mathrm{rest}}\sim f_{\mathrm{min}})}{P_{\rm BA}(\mathbf{k},f_{\mathrm{rest}}\sim f_A)}=1$, and black dashed line indicates $\frac{P_{\rm BA}(\mathbf{k},f_{\mathrm{rest}}\sim f_{\mathrm{min}})}{P_{\rm BA}(\mathbf{k},f_{\mathrm{rest}}\sim f_A)}\propto \chi_{BA}$. Only data with $\chi_{\rm BA}>10^{-2}$ are included to reduce uncertainties. }
 \end{figure}

Based on the analysis of \citet{Zhao2024a}, at perpendicular scales larger than the transition scale ($k_\perp<k_{\rm tran}\sim 5\times10^{-5}$ $\rm km^{-1}$), turbulent energy is concentrated near a characteristic parallel wavenumber $k_{\parallel,0}\sim7\times10^{-5}$ $\mathrm{km^{-1}}$ (Figure~\ref{fig:1D_Alfven}(a)). For $k_\perp>k_{\rm tran}$, $P_{\rm BA}$ follows the anisotropic scaling $k_\parallel\propto k_\perp^{2/3}$, indicating a transition to strong turbulence \citep{Zhao2024a}. We therefore adopt $k_{\parallel,0}$ as a critical threshold to examine how Alfvénic frequency broadening depends on $\chi_{\rm BA}$ (Figure~\ref{fig:Relationship}). The strength of nonlinear frequency broadening is quantified by the power ratio $\frac{P_{\rm BA}(\mathbf{k},f_{\mathrm{rest}}\sim f_{\mathrm{min}})}{P_{\rm BA}(\mathbf{k},f_{\mathrm{rest}}\sim f_A)}$, with smaller values indicating more wave-like (i.e., less broadband) fluctuations.

Figure~\ref{fig:Relationship} reveals two distinct tendencies. (1) For $k_\parallel\leq k_{\parallel,0}$, $P_{\rm BA}$ depends strongly on $k_\perp$ but weakly on $k_\parallel$ (Figure~\ref{fig:1D_Alfven}(a)), indicating a predominantly perpendicular cascade with a quasi-2D structure ($k_\parallel\ll k_\perp$). In this regime, fluctuations reside in a mixed weak–strong regime, containing both wave-like and slowly propagating components (Figure~\ref{fig:1D_Alfven}). The power ratio increases proportionally with $\chi_{\rm BA}$ up to $\chi_{\rm BA}\sim1$, where it approaches unity, then saturates near unity for $\chi_{\rm BA}>1$ (Figure~\ref{fig:Relationship}), indicating a transition from wave-like peaks to frequency-broadened spectra and the onset of strong turbulence \citep{Zhao2024a}. (2) For $k_\parallel> k_{\parallel,0}$, the cascade proceeds to higher $k_\parallel$ and inevitably enters the strong-turbulence regime with a fully three-dimensional cascade. The power ratio remains near unity, indicating the absence of distinct wave–like spectral peaks regardless of the value of $\chi_{\rm BA}$.

 \begin{figure}[t!]
\centering
\includegraphics[scale=0.18]{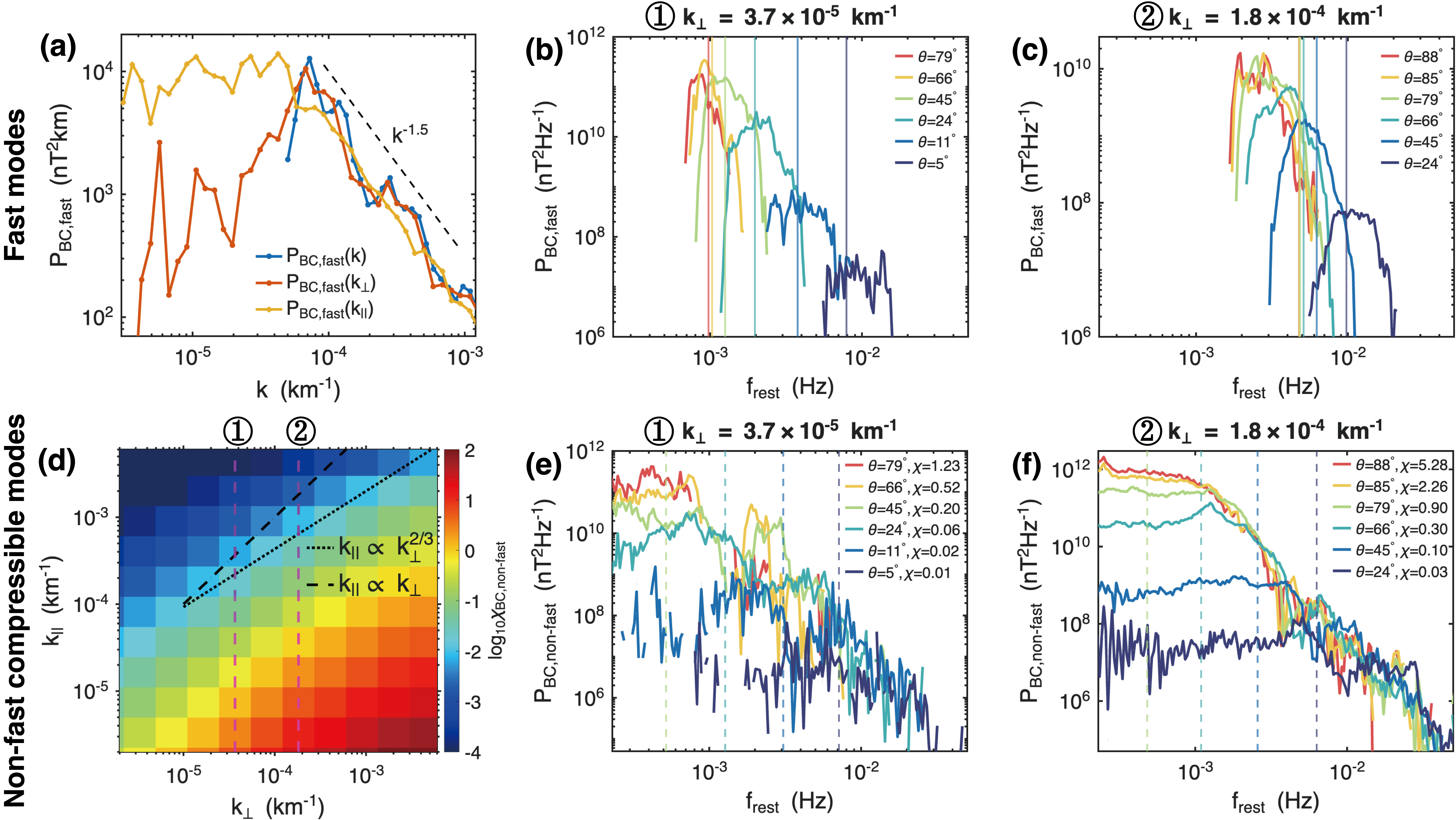}
 \caption{\label{fig:1D_comp} Nonlinear frequency broadening of fast and non-fast compressible magnetic power in the plasma flow frame. (a) 1D fast-mode magnetic energy. (b,c) Fast-mode magnetic power $P_{\rm BC,fast}$ at $k_\perp=3.7\times 10^{-5}~\rm km^{-1}$ and $1.8\times 10^{-4}~\rm km^{-1}$, with color indicating $k_\parallel$. Solid lines represent $f_{\rm fast}$ for the corresponding $k_\parallel$. (d) $k_\parallel-k_\perp$ distribution of the non-fast nonlinearity parameter $\chi_{\rm BC,non\text{-}fast}$. Black dotted (dashed) lines denote the scaling $k_{\rm \parallel} \propto k_\perp^{2/3}$ ($k_{\rm \parallel} \propto k_\perp$), respectively. (e,f) Non-fast compressible magnetic power $P_{\rm BC,non\text{-}fast}$ at $k_\perp=8.2\times 10^{-5}~\rm km^{-1}$ and $1.8\times 10^{-4}~\rm km^{-1}$, with color indicating $k_\parallel$. Dashed lines represent $f_{\rm slow}$ for the corresponding $k_\parallel$. The data are binned into $N_f\times N_{k_\parallel}\times N_{k_\perp}=200\times12\times12$ with $\eta\leq30^\circ$.}
 \end{figure}

 \subsubsection{Compressible modes}

We derive 1D energy spectra of fast modes by integrating the 2D spectrum in Figure~\ref{fig:2D_spec}(g). As shown in Figure~\ref{fig:1D_comp}(a), the spectra exhibit nearly identical power-law scalings: $P_{\rm BC,fast}(k)\propto k^{-1.55}$, $P_{\rm BC,fast}(k_\perp)\propto k_\perp^{-1.57}$, and $P_{\rm BC,fast}(k_\parallel)\propto k_\parallel^{-1.52}$, fitted over $[8\times10^{-5},10^{-3}]~\mathrm{km}^{-1}$. The close agreement further indicates an approximately isotropic distribution of fast-mode energy at these scales. The slightly steeper perpendicular spectrum $P_{\rm BC,fast}(k_\perp)$ relative to the parallel one $P_{\rm BC,fast}(k_\parallel)$ likely reflects collisionless transit-time damping, which preferentially suppresses highly oblique fast modes \citep{Yan2004}, consistent with the depletion of magnetic power near the $\theta=90^\circ$ branch of fast-mode dispersion relations in Figure~\ref{fig:2D_spec}(c).

Figures~\ref{fig:1D_comp}(b,c) show the fast-mode magnetic power spectra $P_{\rm BC,fast}$ at $k_\perp=3.7\times 10^{-5}~\rm km^{-1}$ and $1.8\times 10^{-4}~\rm km^{-1}$. Most spectra exhibit pronounced peaks near the corresponding $f_{\rm fast}$, without low-frequency plateau or high-frequency power-law tails characteristic of strongly nonlinear fluctuations. This behavior indicates that fast modes remain predominantly wave-like, experiencing modest nonlinear frequency broadening, consistent with simulations \citep{Yuen2025}. Consequently, fast-mode fluctuations remain in the weak-turbulence regime and do not transition to strong turbulence as the cascade proceeds to smaller scales.

As $\theta$ approaches $90^\circ$, $P_{\rm BC,fast}$ exhibits enhanced power at frequencies well below $f_{\rm fast}$ (Figures~\ref{fig:1D_comp}(b,c)), likely due to contamination by anti-phase magnetic-density fluctuations near the fast-mode dispersion relations and the limited cascade of fast modes toward higher $k_\perp$ (Figures~\ref{fig:2D_spec}(f,g)). Nevertheless, this effect does not alter the main conclusion that fast modes remain predominantly wave-like and persist in the weak-turbulence regime over the observed wavenumber range.

Non-fast compressible fluctuations are likely dominated by slow modes, which may be passively cascaded by Alfvénic turbulence \citep{Lithwick2001,Cho2003,Schekochihin2009}. By analogy with $\chi_{\rm BA}$, we define the non-fast compressible nonlinearity parameter as $\chi_{\rm BC,non\text{-}fast}=k_\perp\delta \tilde{B}_{\rm BC,non\text{-}fast}/(k_\parallel B_0)$, where the corresponding energy density is $\delta B_{\rm BC,non\text{-}fast}^2(k_\parallel,k_\perp)=\int_0^\infty \int_{k_\parallel}^\infty\int_{k_\perp}^\infty P_{\rm BC,non\text{-}fast}(f_{\mathrm{rest}},k_\parallel,k_\perp)df_{\mathrm{rest}}dk_\parallel dk_\perp$. As shown in Figure~\ref{fig:1D_comp}(d), $\chi_{\rm BC,non\text{-}fast}$ increases either with increasing $k_\perp$ or decreasing $k_\parallel$, indicating stronger nonlinear interactions.

Figures~\ref{fig:1D_comp}(e,f) show the non-fast compressible magnetic power spectra $P_{\rm BC,non\text{-}fast}$ at $k_\perp=3.7\times 10^{-5}~\rm km^{-1}$ and $1.8\times 10^{-4}~\rm km^{-1}$. Similar to Alfvénic fluctuations, for $\chi_{\rm BC,non\text{-}fast}\ll1$, most spectral peaks are concentrated near $f_{\rm slow}$ (Figures~\ref{fig:1D_comp}(e)), confirming that these fluctuations are predominantly wave-like and primarily associated with slow modes. As $\chi_{\rm BC,non\text{-}fast}$ increases, $P_{\rm BC,non\text{-}fast}$ evolves from wave-like peaks to frequency-broadened spectra characterized by a plateau below $f_{\rm slow}$ and a power-law decay at higher frequencies (Figure~\ref{fig:1D_comp}(f)). This evolution provides evidence that slow modes undergo a weak-to-strong turbulence transition, similar to that observed in Alfvénic turbulence \citep{Zhao2024a}.

Overall, the nonlinear frequency broadening of $P_{\rm BC,non\text{-}fast}$ is broadly consistent with that of Alfvénic fluctuations $P_{\rm BA}$, but with broader and less distinct spectral peaks near $f_{\rm slow}$ (Figures~\ref{fig:1D_Alfven} and \ref{fig:1D_comp}). This difference may reflect not only the intrinsically stronger nonlinearity of slow modes \citep{Yuen2025} but also additional non–slow-mode contributions that enhance spectral broadening, such as in-phase magnetic-density fluctuations at ultra-low frequencies (Figure~\ref{fig:2D_spec}(f)). 

\section{Discussion}

We would like to clarify that when $\chi\geq1$ and the spectra become strongly broadened due to nonlinear interactions, the fluctuations may no longer correspond to well-defined linear eigenmodes, and distinct spectral peaks at the corresponding eigenfrequencies may disappear. However, as shown in Figures~\ref{fig:1D_Alfven}(f) and \ref{fig:1D_comp}(f), the eigenfrequencies of the corresponding modes still approximately mark the transition between the spectral plateau and the power-law decay in the power spectra. Therefore, we believe that mode identity remains physically meaningful at MHD scales and is not strictly limited by the local nonlinear parameter. Theoretically, the condition $\chi\sim1$, associated with the onset of strong turbulence \citep{Goldreich1995}, should be understood as an average property of the turbulent cascade rather than a strictly local parameter evaluated at a specific $(k_\perp,k_\parallel)$ location, as considered in this work. In the strong-turbulence regime, the mode properties are more reliably identified through the polarization characteristics of the fluctuations.

In our decomposition framework, MHD mode properties are invoked only to separate compressible and incompressible fluctuations based on their polarization relative to the $\hat{k}\hat{b}_0$ plane. Consequently, slowly propagating compressible fluctuations are included in the compressible-mode category, and the present method does not explicitly distinguish between slowly propagating waves and nonpropagating compressible structures that share the same polarization characteristics. As a result, the non-fast compressible component likely contains contributions from nonpropagating structures and therefore cannot be interpreted exclusively as linear slow modes.

The most direct criterion for distinguishing wave modes from coherent structures is whether the fluctuations propagate in the plasma flow frame. By estimating the phase speed, we find that some fluctuations indeed propagate very slowly. Owing to measurement uncertainties, a strictly zero phase speed in the plasma flow frame cannot be established observationally. Therefore, fluctuations with $V_{\rm ph}\ll V_{\rm A}$ are treated as nonpropagating. For example, the positively correlated fluctuations in the low-frequency range have a phase speed of only $\sim0.2V_{\rm A}$ in the plasma flow frame (Figure~\ref{fig:2D_spec}f). This suggests that at least part of the nonpropagating fluctuations may be embedded in the turbulence associated with the three MHD eigenmodes within our decomposition framework.

One possible interpretation is that, when the nonlinearity becomes sufficiently strong, as shown in Figures~\ref{fig:1D_Alfven}f and \ref{fig:1D_comp}f, the power spectra at fixed $(k_\perp,k_\parallel)$ remain nearly flat below the eigenfrequency, while exhibiting a power-law decay above it. This behavior may suggest that energy begins to cascade only when the fluctuation phase speed exceeds the characteristic phase speed of the corresponding eigenmode. Under this interpretation, fluctuations with frequencies below the eigenfrequency may be approximately regarded as nonpropagating and slowly propagating structures. These fluctuations below the eigenfrequency contain a substantial fraction of the total energy, coincident with the result that quasi-2D magnetic-island-like structures dominate the turbulent energy budget \citep{Gautam2025}. Our preliminary analysis suggests that mirror modes likely dominate the non-propagating fluctuations, and that some of these fluctuations originate from interplanetary magnetic field current sheets. Further categorization is beyond the scope of the present work and will be pursued in future studies.

\section{Conclusion and Implications}

In this study, we advance the polarization-based mode-decomposition method of \citet{Zhao2026}, augmented by incorporating the phase correlation between $\delta |\mathbf{\tilde{B}}|$ and $\delta \tilde{N}$. Using this approach, we obtain spatiotemporal power spectra of the three MHD eigenmodes and present the first quantitative assessment of frequency broadening due to nonlinear interactions. Our main findings are summarized below.

(1) Nonlinear frequency broadening of Alfvén modes exhibits two distinct tendencies at different $k_\parallel$ ranges. (i) For $k_\parallel\leq k_{\parallel,0}$, Alfvén modes evolve from wave-like peaks to frequency-broadened spectra as nonlinearity increases, developing plateaus below $f_{\rm A}$ and power-law tails at higher frequencies, consistent with observations \citep{Zhao2024a}. (ii) In contrast, for $k_\parallel> k_{\parallel,0}$, fluctuations inevitably enter the strong-turbulence regime with a fully three-dimensional cascade, and Alfvénic spectra exhibit no distinct wave-like spectral peak regardless of the value of $\chi_{\rm BA}$ (Figures~\ref{fig:1D_Alfven} and \ref{fig:Relationship}). 

(2) Compressible fluctuations exhibit two distinct components: one follows the fast-mode dispersion relation with in-phase magnetic–density correlations, consistent with fast modes; the other (referred to as non-fast compressible modes) follows the scaling $f_{\mathrm{rest}}\propto k_\perp^{2/3}$ with anti-phase magnetic–density correlations, consistent with slow modes (Figure~\ref{fig:2D_spec}).

(3) Fast-mode spectra retain narrow peaks near $f_{\rm fast}$, indicating weak nonlinearity across the inertial range. In contrast, non-fast compressible modes, dominated by slow modes, evolve from wave-like peaks near $f_{\rm slow}$ to frequency-broadened spectra as nonlinearity increases, indicating that slow modes undergo a transition from weak to strong turbulence (Figure~\ref{fig:1D_comp}).

(4) At large scales ($k_\parallel <10^{-5}$ $\rm km^{-1}$), both $P_{\rm BA}$ and $P_{\rm BC}$ exhibit nearly constant ultra-low-frequency fluctuations around $f_{\rm rest}\sim5\times10^{-4}$ Hz with $k_\parallel\ll k_\perp$ (Figures~\ref{fig:2D_spec}(e) and \ref{fig:v_n_spec}(g)), indicating that both Alfvénic and compressible fluctuations contribute significantly to low-frequency, large-scale quasi-2D magnetic structures.

Overall, these results provide a comprehensive observational characterization of mode-dependent spatiotemporal dynamics in compressible MHD turbulence, with broad implications for turbulence-mediated processes \citep{Schlickeiser2002,Yan2004,YLP2008, Maiti2022}. In particular, we find that slow modes undergo a weak-to-strong transition, whereas fast modes remain weakly turbulent (Figure~\ref{fig:1D_comp}). This extends the weak-to-strong transition well established for Alfvénic turbulence to the compressible regime, providing observational constraints for a unified framework of multiscale, multi-mode dynamics. 

This distinction has direct implications for wave transmission. Ultra-low-frequency (ULF) waves in the solar wind and Earth's foreshock are important drivers of magnetospheric Pc4-5 waves \citep{turc2023transmission,Menk2011}, yet their transmission through the magnetosheath remains unclear \citep{Liu2026}. Our results suggest that fast modes, which remain wave-like with weak frequency broadening (Figure~\ref{fig:1D_comp}), are more likely to reach the magnetopause and drive ULF waves via compressional coupling and field-line resonances. In contrast, Alfvén and slow modes undergo rapid nonlinear decorrelation and are therefore less effective at transmitting coherent wave energy. 

Our results also have implications for turbulent reconnection. Current sheets arise from anisotropic turbulent cascades \citep{Matthaeus1986_MR,Boldyrev2006} and are essential for reconnection \citep{Eyink2015,Lazarian2015}. We find that Alfvén and slow modes dominate the turbulence and preferentially form anisotropic, sheet-like structures, whereas fast modes, with a more isotropic cascade, are less conducive to current-sheet formation. These results establish a mode-dependent link between turbulence and reconnection.

\begin{acknowledgments}

We acknowledge the members of the Cluster spacecraft team and NASA’s Coordinated Data Analysis Web. The Cluster data are available at \url{https://cdaweb.gsfc.nasa.gov}. Data analysis was performed using the IRFU-MATLAB analysis package \citep{Khotyaintsev2024} available at \url{https://github.com/irfu/irfu-matlab}. We acknowledge the use of ChatGPT to improve English grammar and sentence structure. S.Z. acknowledges the support from National Natural Science Foundation of China (NSFC) Excellent Young Scientists Fund (Overseas). T.Z.L. acknowledges the support from National Natural Science Foundation of China (NSFC) Excellent Young Scientists Fund (Overseas). C.H. is supported by the Alexander von Humboldt Foundation.
\end{acknowledgments}

\appendix

\setcounter{figure}{0}

\renewcommand{\thefigure}{A\arabic{figure}}

\renewcommand{\figurename}{Figure}

\section{Overview of the Fluctuations in the Earth's magnetosheath}\label{Appendix:overview}

\begin{figure}[t!]
\centering
\includegraphics[scale=0.3]{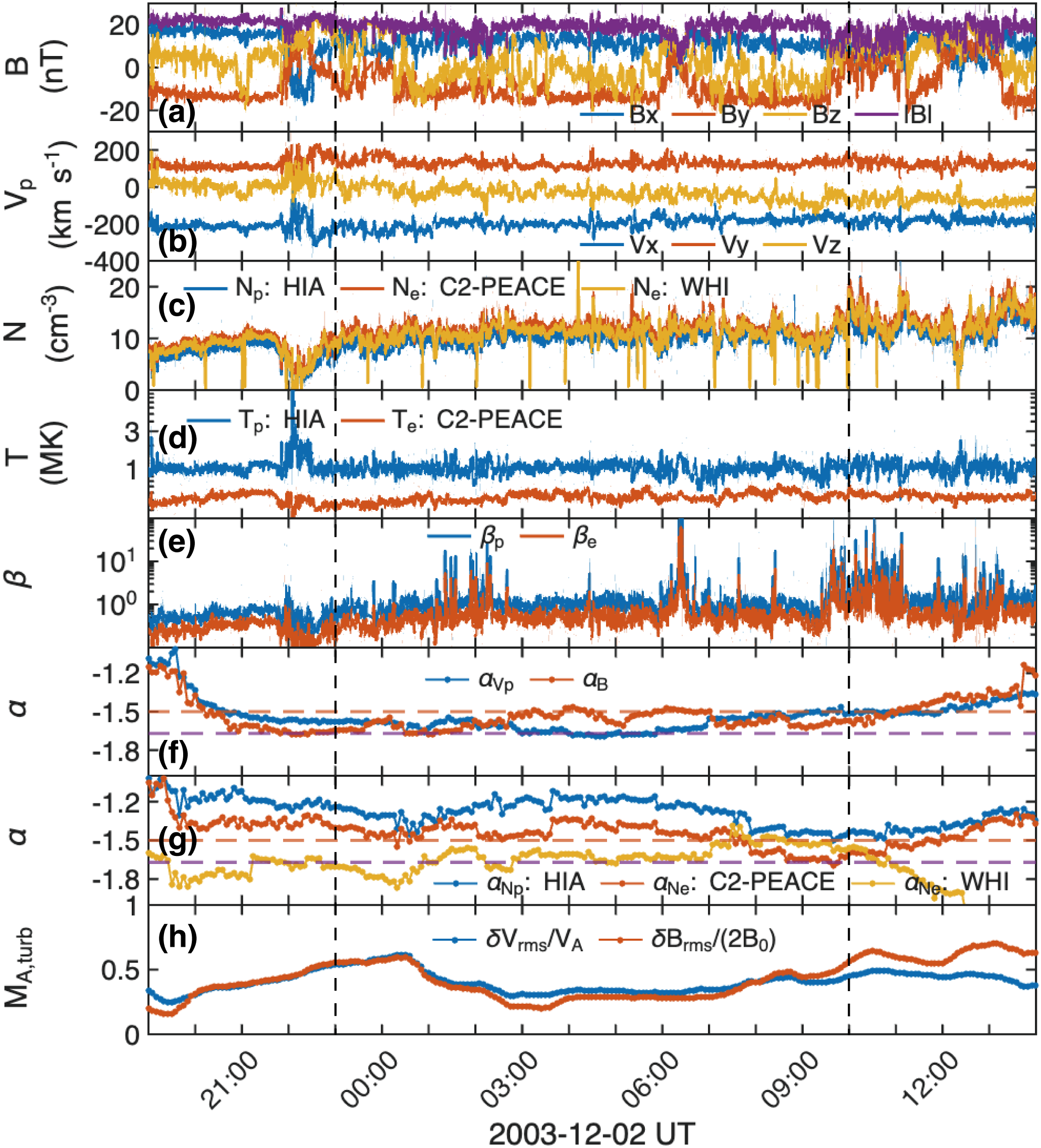}
 \caption{\label{fig:overview} Overview of fluctuations in Earth's magnetosheath in GSE coordinates. (a) Magnetic field. (b) Proton bulk velocity. (c) Proton density from CIS-HIA and electron density from PEACE on Cluster-2 and WHISPER. (d) Temperature. (e) Plasma $\beta$ (the ratio of thermal to magnetic pressures). (f) Spectral slopes ($\alpha$) of trace velocity and magnetic power spectra. (g) Spectral slopes ($\alpha$) of density power spectra. (f,g) The horizontal dashed lines indicate $\alpha=-5/3$ and $-3/2$. (h) Turbulent Alfvén Mach number ($\delta V_{\rm rms}/V_{\rm A}$) and half of the relative amplitudes of the magnetic field ($\delta B_{\rm rms}/(2B_0)$). Unless otherwise noted, all observations are from Cluster-1. The analysis is restricted to the interval between the two vertical dashed lines.}
\end{figure}

Figures~\ref{fig:overview}(a-d) show that the magnetic field and plasma parameters are temporally stable and spatially homogeneous, a conclusion further supported by the time-correlation functions \citep{Zhao2024a}. Figure~\ref{fig:overview}(e) gives the average proton $\beta_{\rm p}\sim1.4$. Figure~\ref{fig:overview}(h) shows $\delta V_{\rm rms}/{V}_{A}<1$ and $\delta B_{\rm rms}/{B}_{0}<1$, confirming the small-amplitude condition. The entire interval is divided into overlapping 5-hour windows with a 5-minute step. For each window, we calculate trace power spectra of velocity, magnetic field, and density fluctuations using a fast Fourier transform with five-point smoothing. Spectral slopes are fitted over $f_{\rm sc}=[0.005f_{\rm cp},0.1f_{\rm cp}]$, where the proton gyrofrequency $f_{\rm cp}\sim0.24$ Hz. Figure~\ref{fig:overview}(f) shows that velocity and magnetic spectra exhibit slopes between $-5/3$ and $-3/2$, indicating well-developed turbulence. Despite discrepancies between proton and electron densities due to instrumental uncertainties (Figure~\ref{fig:overview}(c)), the density spectra remain stable (Figure~\ref{fig:overview}(g)). Moreover, density spectra from both PEACE and WHISPER exhibit slopes close to $-3/2$, indicating the presence of developed compressible turbulence.

\section{Spatiotemporal power spectra of fluctuations}\label{Appen_B}

\begin{figure}[t!]
\centering
\includegraphics[scale=0.14]{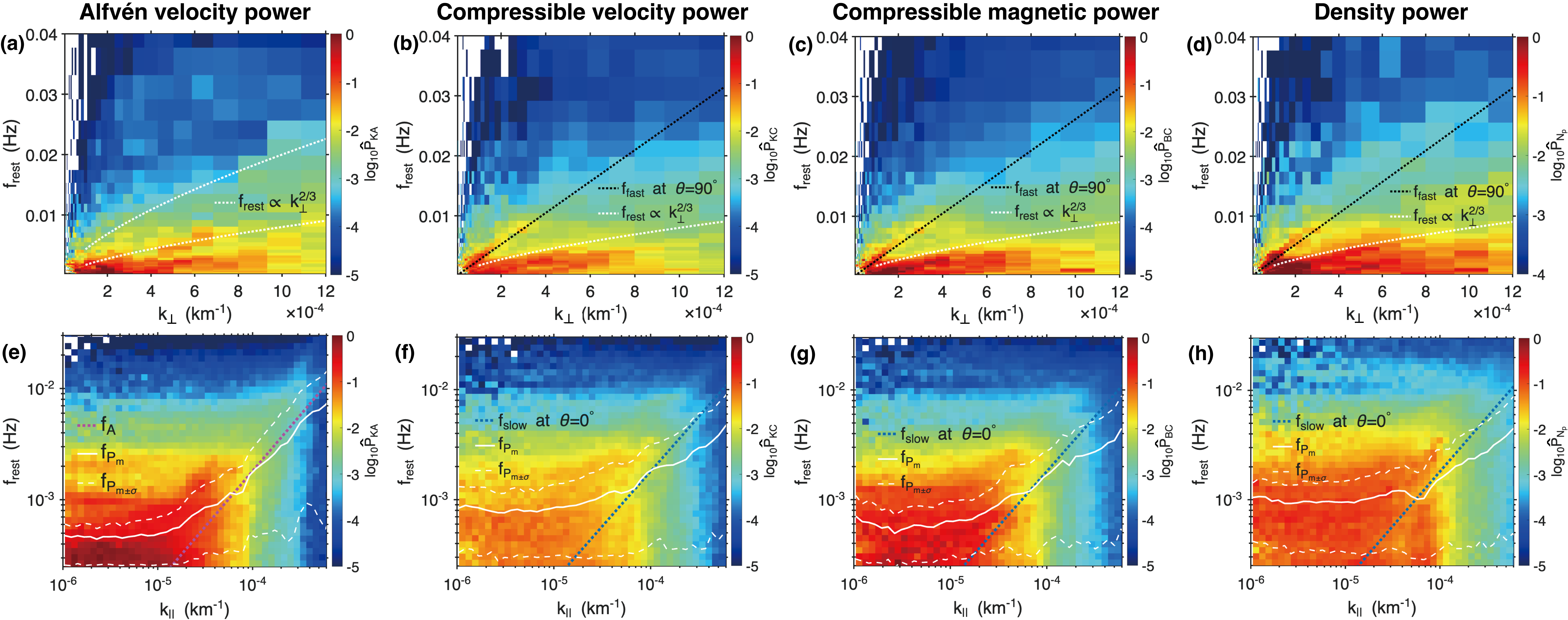}
 \caption{\label{fig:v_n_spec} Spatiotemporal power spectra of velocity and density fluctuations. (a-d) $f_{\mathrm{rest}}-k_\perp$ distributions of Alfvénic velocity power ($\hat{P}_{\rm KA}$), compressible velocity power ($\hat{P}_{\rm KC}$), compressible magnetic power ($\hat{P}_{\rm BC}$), and proton density power ($\hat{P}_{N_p}$). (e-h) $f_{\mathrm{rest}}-k_\parallel$ distributions of $\hat{P}_{\rm KA}$, $\hat{P}_{\rm KC}$, $\hat{P}_{\rm BC}$, and $\hat{P}_{N_p}$. Black dotted lines denote fast-mode dispersion relations at $\theta=90^\circ$. Pink dotted line denotes the dispersion relation of the Alfvén mode. Blue dotted lines denote the dispersion relations of slow modes at $\theta=0^\circ$. White solid curves mark the frequencies of the mean power spectrum, whereas white dashed curves indicate $\pm \sigma$ confidence intervals, where $\sigma$ is the standard deviation of the power. Panels with the same format are normalized by the same factor, and normalized values smaller than $10^{-6}$ are set to NaN. The data are binned into $N_f\times N_{k_\parallel}\times N_{k_\perp}=50\times50\times50$ and restricted to $\eta\leq10^\circ$.}
 \end{figure}

Figure~\ref{fig:v_n_spec} shows the spatiotemporal power spectra of velocity, magnetic, and density fluctuations in the plasma flow frame. The spectra of different physical quantities within the same mode exhibit similar energy distributions, demonstrating the robustness and reliability of the mode decomposition of turbulence.

 \section{Method for Analyzing the Phase Correlation Between Magnetic Field Strength and Plasma Density Fluctuations}\label{Appen_C}
 
The time series of magnetic field strength and plasma density are transformed using the Morlet-wavelet transform \citep{Grinsted2004}, yielding Fourier-space fluctuations in GSE coordinates: $\delta |\mathbf{\tilde{B}}|(t,f_{\rm sc})$ and $\delta \tilde{N}(t,f_{\rm sc})$. The phase difference $\delta\tilde\phi(t,f_{\rm sc})$ is defined as the phase angle of their complex cross-spectrum, calculated by multiplying $\delta |\mathbf{\tilde{B}}|$ with the complex conjugate of $\delta \tilde{N}$. The resulting phase is expressed in radians and lies within the range $(-\pi, \pi]$. 

\begin{figure}[ht!]
\centering
\includegraphics[scale=0.25]{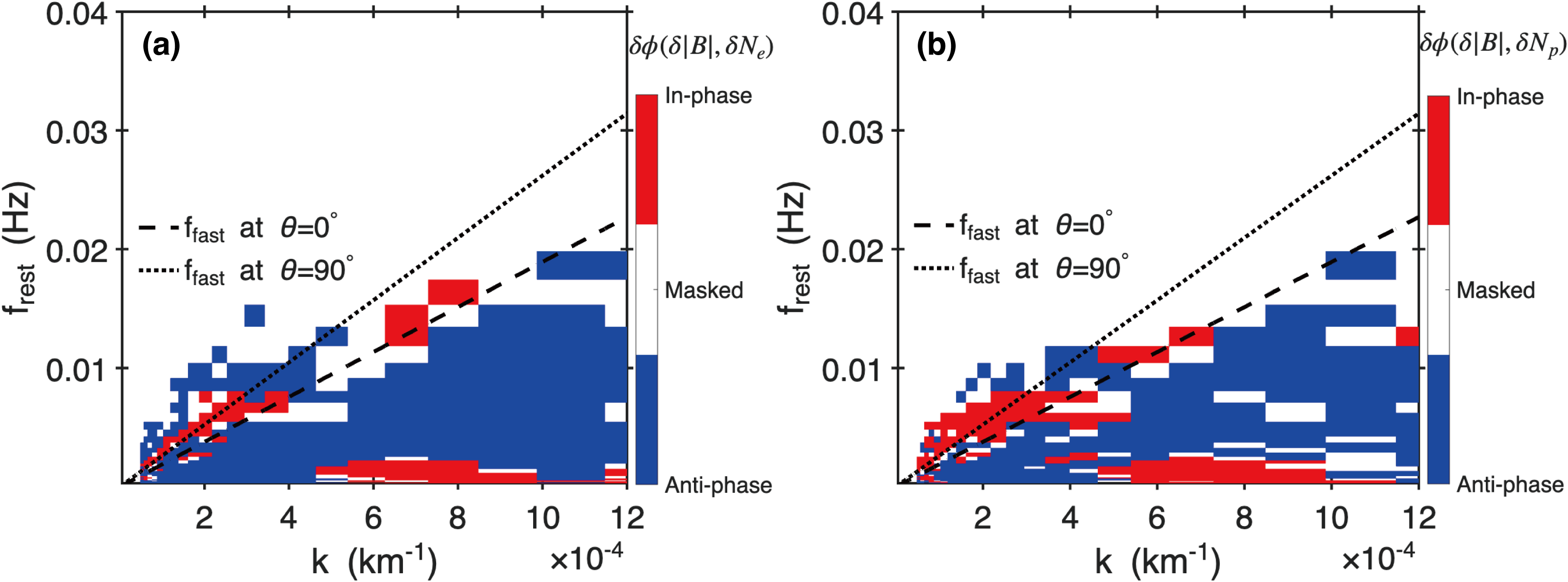}
 \caption{\label{fig:phase_BN} Phase correlation between fluctuations in magnetic field strength and plasma density. (a) Phase correlation between $\delta |\mathbf{\tilde{B}}|$ and CIA-HIA proton density $\delta \tilde{N}_{\rm p}$. (b) Phase correlation between $\delta |\mathbf{\tilde{B}}|$ and  WHISPER electron density $\delta \tilde{N}_{\rm e}$. Black dashed (dotted) lines denote the fast-mode dispersion relations at $\theta=0^\circ$ ($90^\circ$), respectively.}
 \end{figure}

To obtain the spatiotemporal distribution of the phase relation in the plasma flow frame, we construct a set of $N_f\times N_{k_\parallel}\times N_{k_\perp}=50\times 50\times50$ bins in $(f_{\rm rest},k_\parallel,k_\perp)$ space, applying an angle threshold of $\eta\leq10^\circ$. For each time–frequency sample, the phase difference $\tilde\phi$ is represented as a unit complex vector $e^{i\tilde\phi}$ and accumulated with weight $\rm{w}=\sqrt{P_{|B|}P_N}$, where $P_{\rm |B|}=|\delta |\mathbf{\tilde{B}}||^2$ and $P_{\rm N}=|\delta \tilde{N}|^2$. For each $(f_{\rm rest},k_\parallel,k_\perp)$ bin, the weighted circular-averaged phase angle is defined as $\delta\phi = \arg\!\left(\frac{\sum \rm{w}\, e^{i\phi}}{\sum \rm{w}}\right)$, and the corresponding phase-locking value is defined as $L = \left|\frac{\sum \rm{w}\, e^{i\phi}}{\sum \rm{w}}\right|$, which quantifies the coherence of the phase relation within the bin. In-phase fluctuations are identified using a phase-angle tolerance $\rm \pm tol(\delta\phi)$ around $0^\circ$, i.e., $|\delta\phi| \le \rm tol(\delta\phi)$. Anti-phase fluctuations are identified using the same tolerance around $180^\circ$, i.e., $\rm ||\delta\phi|-180^\circ| \le tol(\delta\phi)$. Each bin has an associated energy density. Figures~\ref{fig:2D_spec}(f) and~\ref{fig:phase_BN} display only those bins whose energy density ranks within the top $70\%$ of the cumulative energy and $\rm tol(\delta\phi)=80^\circ$, with the additional requirement $L>0.3$ to ensure phase stability.

\section{Sensitivity of the Results to the Fast-Mode Selection Criteria}\label{Appen_D}

\subsection{Phase-Angle Tolerance Criteria}

\begin{figure}[t!]
\centering
\includegraphics[scale=0.2]{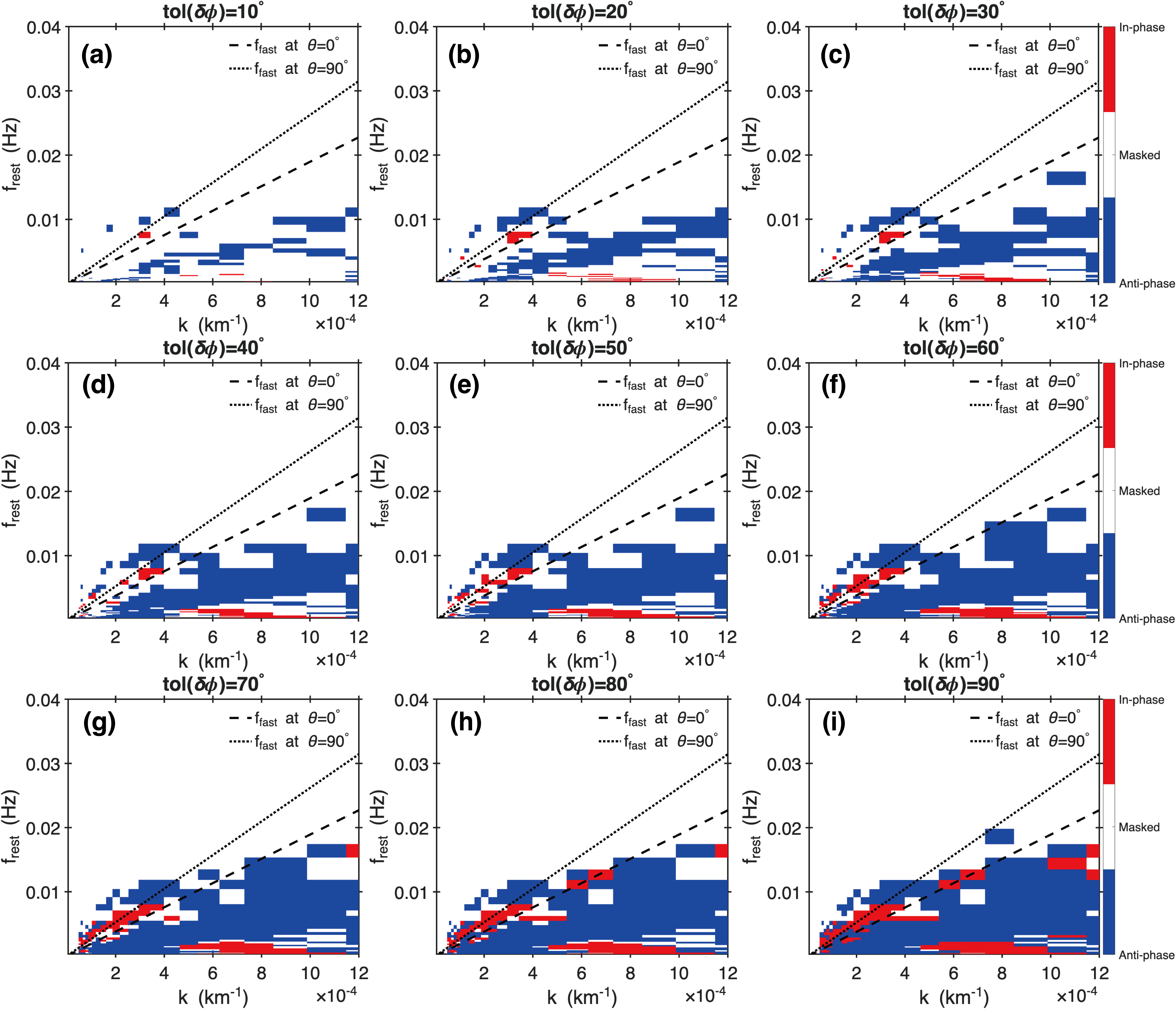}
 \caption{\label{fig:tol_delta_phaseangle} Phase correlation between fluctuations in magnetic field strength ($\delta |\mathbf{\tilde{B}}|$) and PEACE electron density ($\delta \tilde{N}_e$), measured by Cluster-2. In-phase (anti-phase) regions correspond to phase differences within $\pm\rm tol(\delta\phi)$ around $0^\circ$ ($180^\circ$), leaving a transition region between them. Black dashed (dotted) lines denote the fast-mode dispersion relations at $\theta=0^\circ$ ($90^\circ$), respectively. Panels (a–i) correspond to $\mathrm{tol}(\delta\phi)=10^\circ, 20^\circ, \ldots, 90^\circ$, respectively.}
 \end{figure}

Figure~\ref{fig:tol_delta_phaseangle} presents the phase correlation between fluctuations in magnetic field strength ($\delta |\mathbf{\tilde{B}}|$) and PEACE electron density ($\delta \tilde{N}_e$) for different phase-angle tolerance $\rm \pm tol(\delta\phi)$. 

As the phase-angle tolerance criterion is relaxed, an increasing number of bins exhibit stable phase correlations. The stable in-phase fluctuations are consistently observed between the fast-mode dispersion relations at $0^\circ$ and $90^\circ$, as well as in a low-frequency region at $k=[5\times10^{-4}, 10^{-3}]~\mathrm{km}^{-1}$. For $\mathrm{tol}(\delta\phi) \le 70^\circ$, these in-phase correlations near the fast-mode dispersion relation are primarily concentrated at $k < 4\times10^{-4}~\mathrm{km}^{-1}$. As the phase-angle tolerance is further relaxed, the region of stable in-phase correlations near the fast-mode dispersion relation extends to $k \sim 8\times10^{-4}~\mathrm{km}^{-1}$. Overall, the phase correlation between $\delta |\mathbf{\tilde{B}}|$ and $\delta \tilde{N}_e$ remains robust against variations in the phase-angle tolerance criterion.

\subsection{Dispersion-Relation Criteria}

\begin{figure}[t!]
\centering
\includegraphics[scale=0.2]{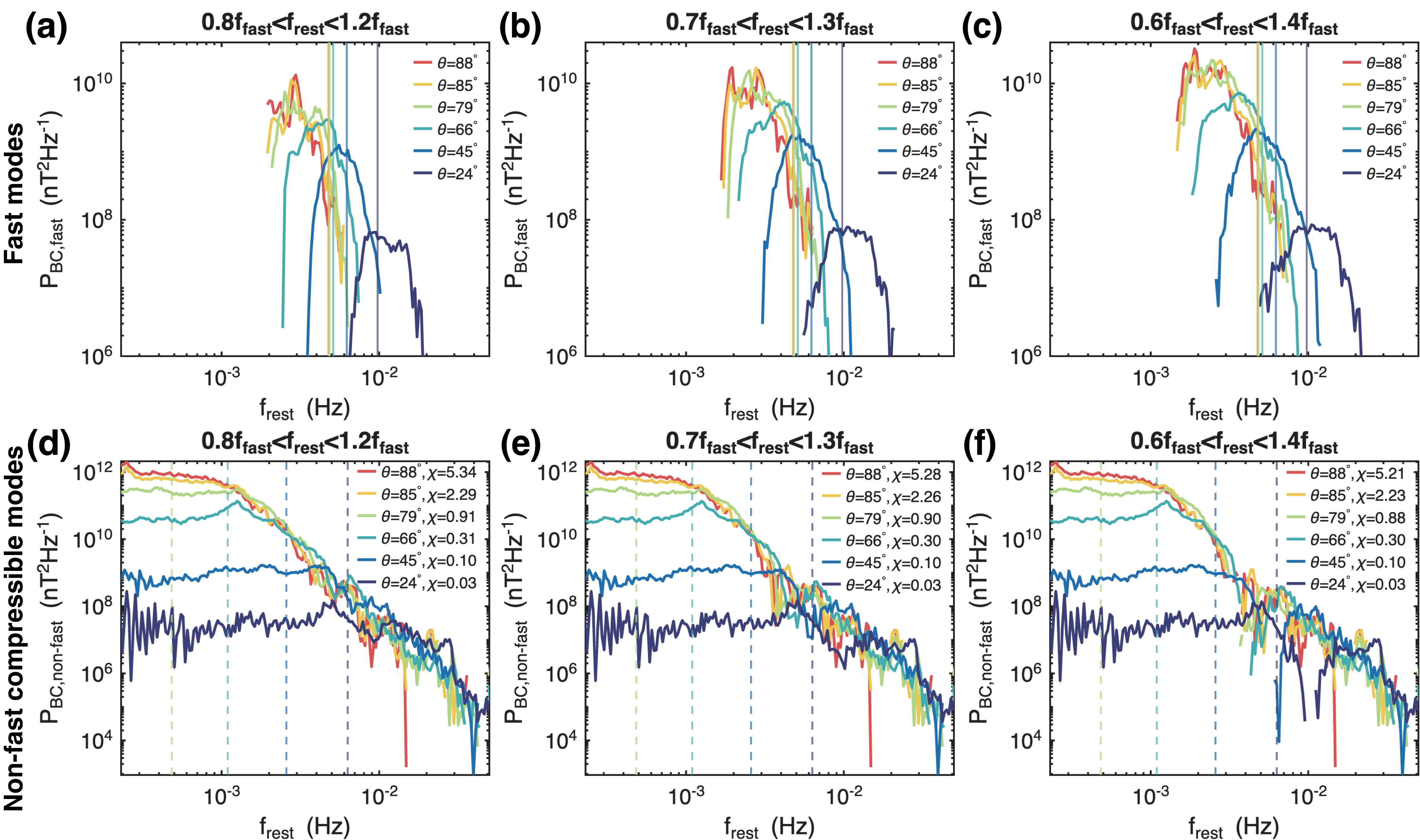}
 \caption{\label{fig:DR_percentage} Nonlinear frequency broadening of fast and non-fast compressible magnetic power in the plasma flow frame at $k_\perp = 1.8\times 10^{-4}~\mathrm{km}^{-1}$, with color indicating $k_\parallel$. Solid and dashed lines represent $f_{\rm fast}$ and $f_{\rm slow}$, respectively, for the corresponding $k_\parallel$. In panels (a,d), (b,e), and (c,f), fast/non-fast components are identified inside/outside $\pm 20\%$, $\pm 30\%$, and $\pm 40\%$ frequency windows centered on $f_{\rm fast}$, respectively. The data are binned into $N_f\times N_{k_\parallel}\times N_{k_\perp}=200\times12\times12$ with $\eta\leq30^\circ$.}
 \end{figure}

Using $\mathrm{tol}(\delta\phi)=80^\circ$ as an example, we further investigate the influence of the dispersion-relation criterion on the inferred weak/strong turbulence transition in fast-mode and non-fast compressible turbulence. Figure~\ref{fig:DR_percentage} shows the nonlinear frequency broadening of fast and non-fast compressible magnetic power in the plasma flow frame for different dispersion-relation criteria. Varying the frequency-window criterion around the fast-mode dispersion relation does not affect our main conclusions. Fast-mode spectra retain narrow peaks near $f_{\rm fast}$ and remain predominantly wave-like, indicating weak nonlinearity across the inertial range. In contrast, non-fast compressible modes, dominated by slow modes, evolve from wave-like peaks near $f_{\rm slow}$ to frequency-broadened spectra as nonlinearity increases, indicating a transition from weak to strong turbulence.

\bibliography{sample701}{}
\bibliographystyle{aasjournalv7}

\end{document}